\documentclass[superscriptaddress,twocolumn,prl]{revtex4-1}
\usepackage{amsmath}
\usepackage{amssymb}
\usepackage{graphicx}
\usepackage{hyperref}
\usepackage{amsfonts}
\usepackage{bbm}
\usepackage{datetime}
\usepackage{siunitx}

\usepackage{color}
\usepackage{xcolor}

\longdate

\renewcommand{\d}{\partial}

\renewcommand{\t}[1]{\tilde{#1}}

\begin{document}
\title{Modeling Leidenfrost levitation of soft elastic solids}
\author{Jack Binysh}
\affiliation{Department of Physics, University of Bath, Claverton Down, Bath BA2 7AY, United Kingdom}
\author{Indrajit Chakraborty}
\affiliation{Mathematics Institute, University of Warwick, Coventry CV4 7AL, United Kingdom}
\author{Mykyta V. Chubynsky}
\affiliation{Mathematics Institute, University of Warwick, Coventry CV4 7AL, United Kingdom}
\author{Vicente Luis D\'{i}az Melian}
\affiliation{IST Austria, Am Campus 1, Klosterneuberg, Austria}
\author{Scott R. Waitukaitis}
\affiliation{IST Austria, Am Campus 1, Klosterneuberg, Austria}
\author{James E. Sprittles}
\affiliation{Mathematics Institute, University of Warwick, Coventry CV4 7AL, United Kingdom}
\author{Anton Souslov}
\email{A.Souslov@bath.ac.uk}
\affiliation{Department of Physics, University of Bath, Claverton Down, Bath BA2 7AY, United Kingdom}

\begin{abstract}
The elastic Leidenfrost effect occurs when a vaporizable soft solid is lowered onto a hot surface. Evaporative flow couples to elastic deformation, giving spontaneous bouncing or steady-state floating. The effect embodies an unexplored interplay between thermodynamics, elasticity, and lubrication: despite being observed, its basic theoretical description remains a challenge. Here, we provide a theory of elastic Leidenfrost floating. As weight increases, a rigid solid sits closer to the hot surface. By contrast, we discover an elasticity-dominated regime where the heavier the solid, the higher it floats. This geometry-governed behavior is reminiscent of the dynamics of large liquid Leidenfrost drops. We show that this elastic regime is characterized by Hertzian behavior of the solid’s underbelly and derive how the float height scales with materials parameters. Introducing a dimensionless elastic Leidenfrost number, we capture the crossover between rigid and Hertzian behavior. Our results provide theoretical underpinning for recent experiments, and point to the design of novel soft machines.
\end{abstract}

\date{\today}
\maketitle
The elastic Leidenfrost effect represents a largely unexplored class of Leidenfrost physics, combining thermodynamics, flow, and elasticity~\cite{waitukaitis_coupling_2017,pham_spontaneous_2017,waitukaitis_bouncing_2018,khattak2019linking,khattak2019microwave}. In the liquid Leidenfrost effect, a fluid droplet hovers above a heated surface, cushioned by a gap layer of its own vapor. The basic physics of this scenario is extensively explored: capillarity and gravity determine the droplet's geometry and how high it floats above the hot surface~\cite{biance_leidenfrost_2003,celestini_take_2012,burton_geometry_2012,quere_leidenfrost_2013, sobac_leidenfrost_2014,sobac_erratum_2021}. These fundamental advances have enabled the discovery of new effects, such as self-propelled droplets~\cite{linke2006self} and controlled wetting~\cite{tran2012drop}, as well as the design of new applications, for example heat exchangers~\cite{van1992physics,vakarelski2012stabilization}.

\begin{figure}[t]
\centering
\includegraphics[width=\linewidth]{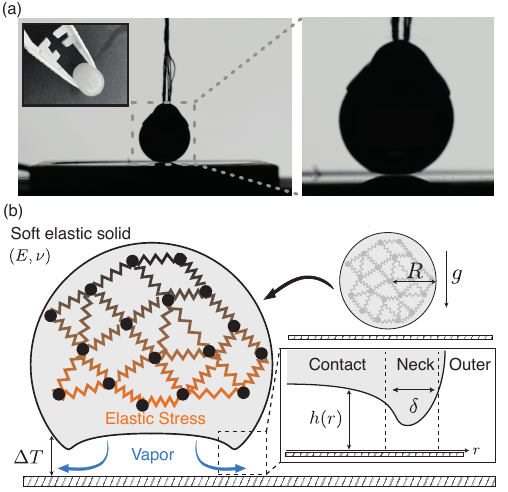}
  \caption{{\bf Leidenfrost levitation of elastic solids enabled by soft lubrication}. (a) 
 A soft elastic hydrogel sphere of radius $R=\SI{7}{\milli\metre}$ hovers above a hot surface ($\Delta T=\SI{115}{\degreeCelsius}$). Inset shows hydrogel in daylight. (b) Evaporative flux elastically deforms the soft solid. Competition between vapor pressure and elastic stress sets the shape of the solid's underbelly and the gap height. {\bf Inset:} We predict distinct height scaling laws in a contact region under the soft solid, an outer region, and a narrow neck region of width $\delta$.}
\label{fig:Setup}
\end{figure}

The typical description of Leidenfrost physics combines flow and phase change, but neglects bulk 3D elastic deformation within the levitated object entirely~\cite{dupeux_viscous_2011,dupeux_self-propelling_2013,wells_sublimation_2015}. Yet, the interplay between fluid flow and soft elastic response is known to yield a plethora of fluid-structure phenomena not possible in a purely rigid limit~\cite{duprat_fluid-structure_2015,gervais_flow_2006,christov_flow_2018,leroy_hydrodynamic_2012, bertin_soft-lubrication_2021, kargar2021lift, hooke_elastohydrodynamic_1972,hamrock_isothermal_1976,archard_non_1968,johnson_regimes_1970,hamrock_fundamentals_2004,skotheim_soft_2004,snoeijer_similarity_2013,essink_regimes_2021,greenwood_elastohydrodynamic_2020}. So it proves in the elastic Leidenfrost effect: when the levitated object is soft and elastic, striking effects result. For example, a water-saturated hydrogel lowered onto a hot surface either bounces spontaneously~\cite{waitukaitis_coupling_2017,pham_spontaneous_2017} or floats on its own vapor layer~\cite{waitukaitis_bouncing_2018}. Figure~\ref{fig:Setup}(a) shows an example of floating behavior for a sphere of radius \SI{7}{\milli\metre}. These effects may appear superficially similar to the phenomenology of liquids~\cite{quere_leidenfrost_2013, graeber_leidenfrost_2021}, but they arise from a distinct interplay between the vapor phase and the condensed phase. In the levitation of Leidenfrost liquids, excess pressure in the vapor layer competes with surface tension~\cite{quere_leidenfrost_2013, sobac_leidenfrost_2014,sobac_erratum_2021, graeber_leidenfrost_2021}. By contrast, in a soft elastic solid [Fig.~\ref{fig:Setup}(a)] the characteristic feature of both bouncing and floating is that the excess pressure in the vapor layer (of order kPa) competes with bulk 3D elastic stress~\cite{waitukaitis_coupling_2017}.

Soft materials thus invite us to re-examine the fundamentals of Leidenfrost physics when combined with large solid-body deformations. However, to fully realize the scope of the elastic Leidenfrost effect, both at a fundamental level and for the potential design of soft devices, a theoretical description of the basic mechanism is required. Despite experimental observation, this description remains a challenge. In particular, there is currently no theory which explains how three-dimensional elasticity determines either the levitation height of the soft solid, or its shape in the floating regime. 

In this Letter, we overcome this challenge by marrying thermodynamic phase change with the lubrication theory of soft elastic objects~\cite{hooke_elastohydrodynamic_1972,hamrock_isothermal_1976,archard_non_1968,johnson_regimes_1970,hamrock_fundamentals_2004,skotheim_soft_2004,snoeijer_similarity_2013,essink_regimes_2021,greenwood_elastohydrodynamic_2020}, to formulate the first description of elastic Leidenfrost floating. By varying a single dimensionless parameter, we discover a transition from rigid behavior to an elasticity-dominated regime described by Hertzian contact mechanics. Using asymptotic analysis and finite element simulations, we quantify this Hertzian limit via scaling laws for the gap height with sphere radius and elastic modulus. Our asymptotic theory reveals the existence of two distinct scalings of the height: the first in a contact region well underneath the solid, and the second in an ever-narrowing neck region [see Fig.~\ref{fig:Setup}(b)]. The development of a neck is also observed for large liquid Leidenfrost drops~\cite{burton_geometry_2012,sobac_leidenfrost_2014,sobac_erratum_2021} and our results invite the question of how liquid Leidenfrost phenomenology intersects with that of soft Leidenfrost solids. More broadly, our results demonstrate how to tailor float height via materials properties, and offer a solid theoretical basis for exploring more complex elastic Leidenfrost phenomena. This theory lays the groundwork for combining elasticity, phase change, and flow to design novel soft machines.

Our first main result is that elastic response yields a new class of scaling laws for the gap height $h$ of floating Leidenfrost objects. This elastic scaling law is distinct from both the liquid and rigid solid cases. A stiff vaporizable sphere (or small liquid drop~\cite{celestini_take_2012}) of radius $R$, density $\rho_s$, and weight $F=(4\pi/3) \rho_s g R^3$ floats at a height $h\sim F^{-1/2} R$ above a heated surface. Taking the load to be proportional to the volume, $F\sim R^3$, we have $h\sim R^{-1/2}$: 
Intuitively, balancing an increasing radius $R$ (i.e., an increasing weight) requires more vapor flux, and so a stiff solid must sit closer to the heated surface. By contrast, we find that a vaporizable \emph{elastic} sphere of Young's modulus $E$ and Poisson ratio $\nu$ (Fig.~\ref{fig:Setup}) has a gap height that scales as 
\begin{equation}
h \sim \Pi_0^{1/4}E^{-1/3}R^{1/3}F^{1/12}.
\label{eq:HertzianHeight}
\end{equation}
In Eq.~\eqref{eq:HertzianHeight}, $\Pi_0$ models the thermal and viscous properties of the vapor layer, and is defined below. Again taking $F\sim R^3$ we find the height scaling $h\sim R^{7/12}$: Counter-intuitively, the heavier the soft solid, the higher it floats.

\begin{figure}[tbp]
\includegraphics[width=0.99\linewidth]{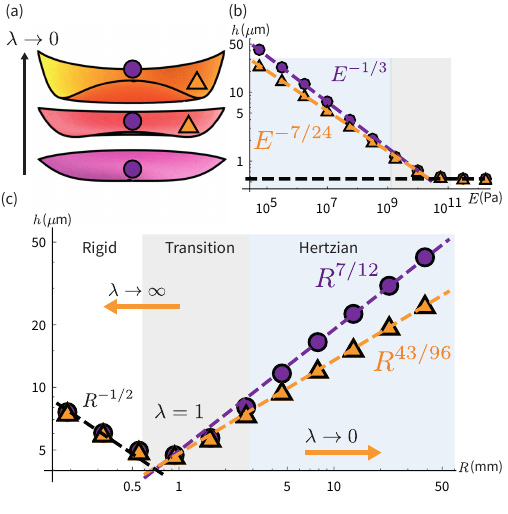}
  \caption{{\bf Gap height scaling laws}. (a) Profiles of the solid's underbelly in the Hertzian limit $\lambda \rightarrow 0$ show the development of a neck region (orange triangle), with height scaling law distinct from the contact region (purple circle). (b--c) Finite element simulations (markers) verify our analytically predicted gap height scaling laws (lines) for the contact ($h \sim E^{-1/3}R^{7/12}$) and neck ($h\sim E^{-7/24}R^{43/96}$) regions. Black lines show analytic predictions for a rigid sphere. We find three regimes: Rigid ($\lambda \rightarrow \infty$), Transition ($\lambda \sim 1$), and Hertzian ($\lambda \rightarrow 0$). In (b), $R=\SI{40}{\milli\metre}$. In (c), $E=\SI{50}{\kilo\pascal}$. Remaining parameters as in~\cite{waitukaitis_bouncing_2018}.} \label{fig:ScalingLaws}
\end{figure}

To derive Eq.~\eqref{eq:HertzianHeight} we now formulate a theory of phase-change induced lubrication coupled to elastic deformation of the solid. Figure~\ref{fig:Setup}(b) shows a schematic of the soft solid floating above a hot surface. The heated surface is held at a temperature difference $\Delta T$ above the vaporization threshold of the solid, causing the solid's underbelly to evaporate and open a thin vapor gap. To describe vapor flow, we note that the gap height is significantly smaller than the lateral scale of the underbelly. We will verify that this observation is indeed self-consistent below. We use the lubrication approximation of the Navier-Stokes equations~\cite{sobac_leidenfrost_2014,hamrock_fundamentals_2004}, which neglects the vertical component of flow. In this approximation, the (axisymmetric) height profile $h(r)$ in Fig.~\ref{fig:Setup}(b) and the pressure in the vapor layer $P(r)$ are related through 
\begin{equation}
\frac{1}{r}\frac{d}{dr} \left(r\frac{\rho h(r)^3}{12 \eta}\frac{dP(r)}{dr}\right) =-\frac{\kappa \Delta T}{L h(r)}.
\label{eq:Reynolds}
\end{equation}
Equation~(\ref{eq:Reynolds}) expresses continuity: the pressure gradient under the solid establishes a Poiseuille flow with mass flux $\sim (\rho/\eta) h^3 \nabla P(r)$, where $\eta$ and $\rho$ are the viscosity and density of the vapor. This flux is balanced by a Leidenfrost source term $-\kappa \Delta T/ L h(r)$, describing conduction-dominated evaporation from the solid's underbelly~\cite{sobac_leidenfrost_2014}.
Here, $\kappa$ is the vapor thermal conductivity and $L$ is the latent heat of vaporization. The materials parameters in Eq.~\eqref{eq:Reynolds} define a typical force scale within the vapor layer, $\Pi_0\equiv\kappa \Delta T \eta/ L\rho$~\cite{dupeux_self-propelling_2013} [see Eq.~\eqref{eq:HertzianHeight}]. Nondimensionalised by the~elastogravitational force scale $E^3/(\rho g)^2$, $\Pi_0$ represents the elastic analog of the evaporation number found in liquid Leidenfrost physics~\cite{sobac_leidenfrost_2014,sobac_erratum_2021}. Using $\Pi_0$, Eq.~\eqref{eq:Reynolds} can be rearranged so that the source term is simply $-\Pi_0/h(r)$.

For a steady gap height, integrated vapor pressure must balance the total weight $F$ of the solid. If the pressure $P$ acts over a lateral length scale $l$ characteristic of the solid's underbelly, we have the scaling $F\sim Pl^2$. A scaling analysis of the lubrication equation Eq.~\eqref{eq:Reynolds} relates $P$, $l$, and gap height $h$ as $P \sim \Pi_0 {l^2}/{h^4}$. Using this pressure relation in the total force balance gives
\begin{equation}
F \sim \Pi_0 \left(\frac{l}{h}\right)^4.
\label{eq:ForceBalance}
\end{equation}
For a given load $F$, Eq.~\eqref{eq:ForceBalance} specifies $h$ in terms of an unknown lateral scale $l$. The crucial question is then: what is the correct choice of $l$? We postulate that there are two choices of $l$, giving two possible gap height scaling laws. The first choice is for a completely rigid sphere, neglecting elasticity: $l_\mathrm{S}=\sqrt{h R}$~\cite{skotheim_soft_2004}. Using this choice in Eq.~\eqref{eq:ForceBalance} recovers the height scaling for rigid spheres, $h\sim\sqrt{\Pi_0/ F} R$. This scaling applies whenever geometric deformation can be neglected~\cite{celestini_take_2012, sobac_erratum_2021}. 

Scaling laws unique to elastic Leidenfrost floating result from a different choice of lateral length scale $l$, arising from linear elasticity theory and Hertzian contact mechanics~\cite{landau_theory_1986,johnson_contact_1985,Bissett1989}: we describe this regime as one of \emph{Hertzian scaling}. When an elastic sphere of Young's modulus $E$ is placed in direct contact with a hard surface, a circular indentation results, with radius $l_H \sim \left(FR/E \right)^{1/3} \sim R^{4/3}$. 
We hypothesize that the underbelly of an elastic Leidenfrost solid asymptotically adopts this lateral scale. The total vapor thrust then scales as the ratio $(l_H/h)^4$, but the total load scales as the volume $R^3$, resulting in a float height given by $h\sim l_H/R^{3/4} \sim R^{7/12}$. Note that $h/l_H \sim R^{-3/4}$, and so the lubrication approximation improves as we go further into the Hertzian limit.

The full scaling with all materials parameters is given in Eq.~\eqref{eq:HertzianHeight}. Intuitively, as the sphere radius increases, elastic deformation of the solid's underbelly gives a rapidly increasing contact area over which evaporative thrust is generated. This increasing thrust outcompetes the increasing weight, leading to the counter-intuitive increase of gap height with radius $R$. In the discussion, we compare this behavior to that of large liquid Leidenfrost drops, which also exhibit a regime of increasing gap height with lateral extent~\cite{biance_leidenfrost_2003, burton_geometry_2012,sobac_leidenfrost_2014,sobac_erratum_2021}.

We have described two distinct scaling regimes for the gap height of elastic Leidenfrost solids: a stiff regime characterized by the lateral length scale $l_{\mathrm{S}}$, and a Hertzian regime characterized by $l_H$. Our second main result is to show that the crossover between these regimes is characterized by a single dimensionless \emph{elastic Leidenfrost number} $\lambda$, defined as 
\begin{equation}
\begin{aligned}
\lambda &\equiv \frac{2\pi}{3} \left[\frac{l_\mathrm{S}}{l_H}\right]^4= \frac{2\pi}{3} \left[\frac{4 E}{3(1-\nu^2)}\right]^{4/3}\Pi_0\, F^{-7/3} R^{8/3}.
\label{eq:lambda}
\end{aligned}
\end{equation}
Intuitively, $\lambda$ compares the length scales over which vapor pressure causes elastic deformation, as shown by the first equality in Eq.~\eqref{eq:lambda}. The second equality provides an expression in terms of materials parameters.
 When $\lambda \rightarrow \infty$, $l_{\mathrm{S}}\gg l_H$ and vapor pressure is too small to cause appreciable elastic deformation. By contrast, when $\lambda \rightarrow 0$, $ l_\mathrm{S} \ll l_H$ and Hertzian elasticity dominates. A crossover between the rigid and Hertzian regimes is expected at $\lambda \sim 1$. In the SM~\cite{SI}, we show that non-dimensionalizing the combined equations of linear elasticity and the lubrication equation [i.e., Eq.~\eqref{eq:Reynolds}] yields $\lambda$ as the single dimensionless number governing the floating regime. 
 
We have predicted that the dimensionless parameter $\lambda$ mediates the crossover between rigid behavior and our scaling law, Eq.~\eqref{eq:HertzianHeight}. We now test these predictions. To do so, we numerically solve for a series of profiles for the gap height $h(r)$ and for the pressure $P(r)$, across a range of sphere radii and Young's moduli. We implement a hybrid finite element method in COMSOL Multiphysics, in which the equations of linear elasticity are solved throughout the 3D solid. This elastic solver is coupled to a numerical solution of the lubrication equation Eq.~\eqref{eq:Reynolds} via COMSOL's standard \texttt{Coefficient Form Boundary PDE} option. Our finite element approach, described further in the SM~\cite{SI}, was used in Refs.~\cite{chubynsky_bouncing_2020, chakraborty_computational_2022} to study droplet impact and the liquid Leidenfrost effect. This method allows us to probe the limits of validity for our theory by bypassing the assumptions made in Hertzian contact theory, i.e., the use of a half-space elastic solution for a curved boundary and a parabolic approximation to the solid's underbelly. 

 In Fig.~\ref{fig:ScalingLaws}, we show the gap height in the contact region, $h(r=0)$, against radius $R$ and modulus $E$. Parameters not varied are fixed to natural experimental values for the hydrogel spheres used in, for example, Ref.~\cite{waitukaitis_bouncing_2018}. We find a clear crossover between two distinct regimes of behavior occurring at $\lambda \sim 1$, with agreement between our predicted scaling laws, Eq.~\eqref{eq:HertzianHeight}, and those found in simulation. However, our numerical results also reveal a neck region at the edge of contact [Fig.~\ref{fig:ScalingLaws}(a)], which develops as the solid transitions into the Hertzian regime. The height of this neck follows a distinct scaling law, not captured by the analysis above.
 
\begin{figure}[htbp]
\centering
\includegraphics{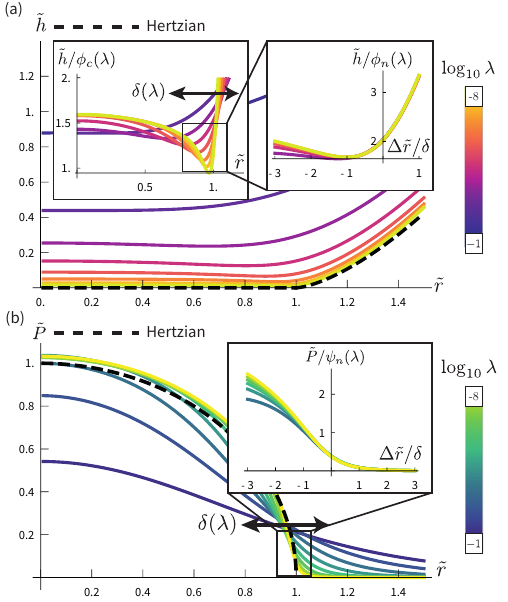}
  \caption{{\bf Collapsing to the Hertzian Limit.} Nondimensionalized (a) height $\t h$ and (b) pressure $\t P$ profiles from finite element simulation. Both approach the Hertzian solutions (black dashed lines) as $\lambda \rightarrow 0$. Deviations are confined to the neck region $\delta(\lambda)$. {{\bf Insets}}: Our height scaling law in the contact region, $\phi_c(\lambda)=\lambda^{1/4}$, breaks down in the neck (a, left). Instead, our asymptotic theory predicts that profiles collapse in the neck when radius $\Delta \t r \equiv \t r -1$ is rescaled by $\delta(\lambda)=\lambda^{3/16}$, height by $\phi_n(\lambda)=\lambda^{9/32}$ (a, right) and pressure by $\psi_n(\lambda)=\lambda^{3/32}$ (b, right).}
\label{fig:CollapsingHertzian}
\end{figure}
To study this neck region further, in Fig.~\ref{fig:CollapsingHertzian} we plot the full height [Fig.~\ref{fig:CollapsingHertzian}(a)] and pressure [Fig.~\ref{fig:CollapsingHertzian}(b)] profiles under the soft solid, non-dimensionalized by Hertzian scales: $\t r = r/l_H$, $\t h = h R/l^2_H$, $\t P =(2\pi l_H^2/3F) P$. As $\lambda\rightarrow0$ both height and pressure profiles approach their Hertzian limits, $\t h (\t r)=(\t r-1)^{3/2}$ for $\t r \gtrsim 1$, and $\t P(\t r)= \sqrt{1- \t r^2}$ for $\t r<1$~\cite{johnson_contact_1985}, except in a boundary layer of width $\delta(\lambda)$ located at $\t r=1$. The discrepancy in the height data becomes clearer when we rescale $\t h$ by the contact scaling law Eq.~\eqref{eq:HertzianHeight}. We show in the SM~\cite{SI} that Eq.~\eqref{eq:HertzianHeight} corresponds to the dimensionless scaling law $\t h(\t r=0) \sim \phi_c(\lambda)$, where $\phi_c(\lambda)=\lambda^{1/4}$. As shown in the left inset of Fig.~\ref{fig:CollapsingHertzian}(a), this law collapses data in the contact region, but fails in the neck $\delta(\lambda)$. The reason is that the Hertzian dry contact solutions are singular at $\t r =1 $. This singularity implies a breakdown of Hertz theory over the width $\delta(\lambda)$, because the height and pressure profiles in our lubrication problem must remain smooth everywhere. In this region, the height scaling from Eq.~\eqref{eq:HertzianHeight} does not apply because the relevant lateral length scale is no longer the Hertzian length scale $l_H$.

To capture the anomalous scaling of the height in the neck region and the width $\delta(\lambda)$, we take inspiration from the numerical collapse of Fig.~\ref{fig:CollapsingHertzian}. The key observation is that in the contact region under the solid ($\t r \ll 1$), the pressure is given by the Hertzian solution at leading order in the parameter $\lambda$ [Fig.~\ref{fig:CollapsingHertzian}(b)]. By the same logic, when $\t r \gg 1$, the height is asymptotically Hertzian [Fig.~\ref{fig:CollapsingHertzian}(a)]. Using the lubrication equation Eq.~\eqref{eq:Reynolds}, we construct the corresponding height and pressure solutions in each region. These solutions patch together over the neck region, shown schematically in the inset of Fig.~\ref{fig:Setup}(b). In the neck, both pressure and height vanish as some unknown power of $\lambda$; we denote the height scaling as ${\phi_n(\lambda)}$ and the pressure scaling as $\psi_n(\lambda)$. The patching conditions, derived in the SM~\cite{SI}, determine $\delta(\lambda)$, $\phi_n(\lambda)$, and $\psi_n(\lambda)$ to give a complete set of scaling laws:
\begin{equation}
\begin{aligned}
\delta(\lambda)=\lambda^{3/16}, \ \phi_c(\lambda)=\lambda^{1/4}, \\ 
 \psi_n(\lambda)=\lambda^{3/32}, \ \phi_n(\lambda)=\lambda^{9/32}. 
\end{aligned}\label{eq:ScalingLawsNonDim}
\end{equation}
In the insets of Fig.~\ref{fig:CollapsingHertzian}, we show that the scalings Eq.~\eqref{eq:ScalingLawsNonDim} now collapse our simulation data in the neck region as well as the contact region. Our asymptotic theory gives a new prediction: re-dimensionalized, the relation $\phi_n(\lambda)=\lambda^{9/32}$ yields the anomalous neck height scaling
\begin{equation}
h \sim \Pi_0^{9/32}E^{-7/24}F^{1/96}R^{5/12}.
\label{eq:NeckHeight}
\end{equation}
Again taking the load to go as the volume, $F\sim R^3$, we find the neck height scaling $h\sim R^{43/96}$. In Fig.~\ref{fig:ScalingLaws}, we show that these revised scalings with radius $R$ and modulus $E$ agree well with simulations. Taken together, the scalings Eqs.~\eqref{eq:HertzianHeight} and~\eqref{eq:NeckHeight} provide a complete picture of elastic Leidenfrost floating, with the agreement between the asymptotic result Eq.~\eqref{eq:NeckHeight} and our simulations also serving as a rigorous cross-check on our theory.

Our fundamental description of elastic Leidenfrost floating provides the theoretical groundwork for interpreting recent studies~\cite{waitukaitis_coupling_2017, waitukaitis_bouncing_2018}, and establishes principles for experimental investigation of this new class of Leidenfrost phenomena.
Using hydrogel spheres of radius $R=\SI{7}{\milli\metre}$ and modulus $E=\SI{50}{\kilo\pascal}$, Ref.~\cite{waitukaitis_bouncing_2018} places an upper bound on the gap height in the floating regime as $h < \SI{25\pm10}{\micro\metre}$. Our theory predicts a contact height of $h=\SI{15}{\micro\metre}$ and a neck height of $h=\SI{12}{\micro\metre}$, and finds $\lambda\sim 10^{-5}$, placing the experiments of Ref.~\cite{waitukaitis_bouncing_2018} in the regime of Hertzian scaling governed by Eq.~\eqref{eq:HertzianHeight}. Gap heights of $\sim\SI{15}{\micro\metre}$ are measurable via interferometric imaging, although inferring absolute height data in this range requires techniques beyond white-light interferometry: In the SM we describe the experimental methodology necessary to probe our theoretical scaling laws. 

Before chimneying, large liquid Leidenfrost drops also exhibit a regime of increasing float height with lateral extent, and the development of a neck~\cite{biance_leidenfrost_2003, burton_geometry_2012,sobac_leidenfrost_2014,sobac_erratum_2021}. The mechanism behind this regime, both in liquids and the soft elastic solids considered here, is geometric change occurring on the underbelly of the levitated object. However, scaling relations differ between the liquid and soft solid cases~\cite{sobac_leidenfrost_2014}. For example, we find neck height scaling $h \sim R^{43/96}$, whereas in Ref.~\cite{sobac_erratum_2021} the neck height appears to plateau at a constant value. Our work invites the question of how much of the rich phenomenology of liquids finds an elastic counterpart~\cite{bouillant_leidenfrost_2018,bouillant_self-excitation_2021}.

More broadly, our work points towards combining Leidenfrost-type physics and soft elasticity beyond the setup of Fig.~\ref{fig:Setup}(a). We envision tailoring the floating configuration of an object by combining phase-change induced forces with those from motion~\cite{skotheim_soft_2004,essink_regimes_2021}, and by tuning initial geometry: In the SM~\cite{SI}, we show that an elastic cylinder in the Hertzian regime has a contact height scaling $h\sim R^{5/8}$, distinct from the spherical case. Such shape control is not possible for liquid droplets.

\begin{acknowledgments}{We are grateful to Dominic Vella, Jens Eggers, John Kolinski, Joshua Dijksman, and Daniel Bonn for insightful discussions. The supporting data for this article are openly available from Zenodo at DOI: 10.5281/zenodo.8329176 under an MIT license. J.B.~and A.S.~acknowledge the support of the Engineering and Physical Sciences Research Council (EPSRC) through New Investigator Award No.~EP/T000961/1. A.S.~acknowledges the support of Royal Society under grant No.~RGS/R2/202135. J.E.S.~acknowledges EPSRC Grants No.~EP/N016602/1, EP/S022848/1, EP/S029966/1, and EP/P031684/1.}
\end{acknowledgments}

%


\makeatletter
\makeatother

\renewcommand{\d}{\partial}
\renewcommand{\t}[1]{\tilde{#1}}

\onecolumngrid
\pagebreak
\setcounter{equation}{0}
\renewcommand{\theequation}{S\arabic{equation}}
\renewcommand{\thefigure}{S\arabic{figure}}
\setcounter{figure}{0}
\setcounter{secnumdepth}{3}

\section*{Supplementary Material}

\section{Introduction}
In this Supplementary Material, in \S\ref{sec:Formulate} we formulate the coupled equations of fluid flow and linear elasticity which describe the underbelly of the soft solid. Non-dimensionalizing this system by Hertzian scales provides a natural definition of the elastic Leidenfrost number $\lambda$, Eq.~(4) of the main text. In \S\ref{sec:Asymptotic}, we detail our asymptotic analysis in the limit $\lambda \rightarrow 0$. In \S\ref{sec:ScalingLaws}, we use our asymptotics to derive the scaling laws for the contact and neck height, Eqs.~(1) and (6) of the main text, respectively. In \S\ref{sec:Experiment} we describe the interferometric techniques necessary for probing our theoretically derived scaling laws, and give a statistical analysis of the required experimental measurement accuracy. In \S\ref{sec:Numerics}, we describe our finite element simulation method, and in \S\ref{sec:Deviations}, we show that deviations from Hertzian predictions vanish in the limit $l_H/R \rightarrow 0$. In \S\ref{sec:Cylinder}, we derive the scaling law for the gap height of a cylinder. Finally, in \S\ref{sec:Materials}, for the reader's convenience we reproduce the hydrogel materials parameters found in Ref.~\cite{waitukaitis_bouncing_2018}.

\section{Formulating the elastohydrodynamic equations}
\label{sec:Formulate}

\begin{table}[h]
  \begin{center}
    \begin{ruledtabular}
    \begin{tabular}{ccc} 
      Name & Symbol & Definition\\ 
      \hline
      \textbf{Spherical Geometry} \\
      \hline
      Sphere Radius & $R$ & \\
      Sphere Density & $\rho_s$ & \\
      Sphere Weight & F & $\frac{4\pi}{3} \rho_l g R^3$ \\
      \hline
      \textbf{Elasticity} \\
      \hline
      Young's Modulus & E & \\
      Poisson Ratio & $\nu$ & \\
      Hertzian Contact Radius &$l_H$ & $\left[\frac{3(1-\nu^2)}{4}\frac{FR}{E}\right]^\frac{1}{3}$\\
      Hertzian Vertical Deformation & $\delta_H$ & $\frac{l_H^2}{R}$\\
      Hertzian Pressure & $P_H$ & $\frac{3F}{2\pi l_H^2}$ \\
      \hline
      \textbf{Lubrication Theory} \\
      \hline
      Temperature gap & $\Delta T$ \\
      Thermal Conductivity & $\kappa$ \\
      Latent Heat of Vaporization & $L$ \\
      Viscosity & $\eta$ \\
      Vapor Density & $\rho$ \\
      Fluid Flux Scale & $ \Pi_0$ & $\kappa \Delta T \eta/ L\rho$ \\
    \end{tabular}
    \end{ruledtabular}
  \end{center}
    \caption{Definitions of quantities used throughout the manuscript. Elasticity theory definitions are consistent with those found in~\cite{landau_theory_1986,johnson_contact_1985}.} 
    \label{tab:scales}
\end{table}

Our starting point is the coupled equations of linear elastic deformation and lubrication theory. Fluid flow is described by the lubrication equation, Eq.~(2) of the main text. For elastic deformations, in Cartesian coordinates $(x,y)$, the vertical deflection $u(x,y)$ of a half-space due to an applied pressure field $P(x,y)$ is given using the linear-elastic Green's function by~\cite{landau_theory_1986,johnson_contact_1985}
\begin{equation}
u(x,y) = \frac{1-\nu^2}{\pi E} \int \frac{P(x',y')}{\sqrt{(x-x')^2+(y-y')^2}} d{x'}d{y'}.
\label{eq:VerticalDeflection}
\end{equation}
In the case of an axisymmetric pressure profile $P(r)$ (where $r=\sqrt{x^2+y^2}$), Eq.~\eqref{eq:VerticalDeflection} simplifies to
\begin{equation}
  u(r)=\frac{4(1-\nu^2)}{\pi E} \int P(r')\frac{r'}{r+r'} K\left(\sqrt{\frac{4rr'}{(r+r')^2}}\right) dr',
  \label{eq:SymmetricVerticalDeflection}
\end{equation}
where $K(k)$ is the complete elliptic integral of the first kind with modulus $k$~\cite{johnson_contact_1985}. We approximate the deformation of the solid's underbelly by the half-space expression Eq.~\eqref{eq:SymmetricVerticalDeflection}. The height profile $h(r)$ is given by an initial, undeflected profile, plus the elastic deflection $u(r)$. For the initial profile, we make a parabolic approximation to a sphere. The full height profile is then
\begin{equation}
h(r)=h_0 + \frac{r^2}{2 R} + u(r).
\label{eq:HeightProfile}
\end{equation}
In Eq.~\eqref{eq:HeightProfile}, we have an undetermined constant $h_0$, which will be set by the global constraint of total force balance. 

In summary, our coupled system of elastic deformations and fluid flow reads 
\begin{equation}
  \begin{aligned}
  \frac{1}{r}\d_r \left(r\frac{h^3}{12}\d_r P \right)
  =-\frac{\Pi_0}{h}, \\
  h(r)=h_0 + \frac{r^2}{2R} +\frac{4(1-\nu^2)}{\pi E} \int P(r')\frac{r'}{r+r'} K\left(\sqrt{\frac{4rr'}{(r+r')^2}}\right) dr', \\
  2\pi \int r P(r)dr = F.
  \end{aligned}
  \label{eq:System}
\end{equation}
Equation~\eqref{eq:System} is the system from which we derive $\lambda$, and scaling laws for the gap height. We note that it is a closed system of three equations in three unknowns: $h_0, h(r), P(r)$. For ease of reference, Table~\ref{tab:scales} summarizes the notation used for materials parameters and variables in this section and throughout this Supplemental Material.  

\subsection{Non-dimensionalization}
We now non-dimensionalize Eq.~\eqref{eq:System} using Hertzian scales:
\begin{equation}
\begin{aligned}
  \tilde{r}= r/l_H, \\
  \tilde{h} = h/\delta_H, \\
  \tilde{P}= P/P_H.
  \label{eq:HertzianNonDim}
\end{aligned}
\end{equation}
Here, $l_H$ is the Hertzian contact radius, $\delta_H= l_H^2/R$ is the typical scale of vertical deflection in Hertzian contact mechanics, with $P_H=3 F/2\pi l_H^2$ the typical pressure scale. We summarize these definitions in Table~\ref{tab:scales}. Exact Hertzian definitions are chosen to be consistent with those found in, e.g., Refs.~\cite{johnson_contact_1985, landau_theory_1986}. Substituting the definitions in Eq.~\eqref{eq:HertzianNonDim} into our coupled system, Eq.~\eqref{eq:System}, yields
\begin{equation}
  \begin{aligned}
    \frac{1}{\t r}\d_{\t r} \left(\t r\frac{ \t h^3}{12}\d_{\t r} \t P \right) =-\frac{\lambda}{\t h}, \\
    \t h(\t r)=\t h_0 + \frac{\t r^2}{2} +\frac{8}{\pi^2} \int \t P(\t r')\frac{\t r'}{\t r+\t r'} K\left(\sqrt{\frac{4\t r\t r'}{(\t r+\t r')^2}}\right) d \t r', \\
    \int \t r \t P(\t r)d \t r = \frac{1}{3}. 
  \label{eq:SystemNonDim}
  \end{aligned}
\end{equation}
Non-dimensionalized by Hertzian scales, the elastic Leidenfrost number $\lambda$ naturally appears as the sole parameter in our system, Eq.~\eqref{eq:SystemNonDim}. Following the approach above, an initial expression for $\lambda$ is given by
\begin{equation}
\lambda \equiv \frac{\Pi_0 l^2_H}{P_H \delta^4_H}.
\label{eq:lam1}
\end{equation}
However, there are several ways to re-express $\lambda$ that we find more insightful. 

\subsection{Interpretations of the elastic Leidenfrost number}
We now give several expressions for the elastic Leidenfrost number $\lambda$, which highlight its conceptual usefulness. First, in terms of materials parameters, we find 
\begin{equation}
\lambda = \frac{2\pi}{3}\left[\frac{4}{3(1-\nu^2)}\right]^{4/3} \Pi_0 F^{-7/3} R^{8/3} E^{4/3},
\label{eq:lammaterials}
\end{equation}
an expression given in Eq.~(4) of the main text. Equation~\eqref{eq:lam1} can be rewritten in terms of the Hertzian contact radius $l_H$ and the stiff sphere length scale (discussed in the main text), defined as $l_S \equiv \sqrt{hR}$ and given by $l_S\equiv (\Pi_0/F)^{1/4} R$ via Eq. (3). Expressing $\lambda$ in terms of $l_H$ and $l_S$ gives
\begin{equation}
\lambda = \frac{2 \pi}{3} \left(\frac{l_S}{l_H}\right)^4.
\label{eq:lamlength}
\end{equation}
Equation~\eqref{eq:lamlength} expresses $\lambda$ as a crossover between length scales. The elastic Leidenfrost number $\lambda$ can also be interpreted as a crossover of \emph{pressure} scales~\cite{johnson_regimes_1970}. Taking the vapor pressure scale under a stiff solid as $P_S=F/l_S^2$, we obtain
\begin{equation}
\lambda = \left(\frac{2 \pi}{3}\right)^3 \left(\frac{P_H}{P_S}\right)^2.
\label{eq:lamPressure}
\end{equation}
The intepretation of Eq.~\eqref{eq:lamPressure} is that, when $\lambda \rightarrow \infty$, $P_S \ll P_H$, and the pressure scale under the soft solid is much smaller than the Hertzian pressure scale. In this limit, we do not expect substantial elastic deformation. In the opposite limit, $\lambda \rightarrow 0$, $P_S \gg P_H$ and the pressure scale under a hypothetically stiff sphere is far greater than the Hertzian pressure. In this limit, we expect large elastic deformation. A final useful expression for $\lambda$, in terms of a ratio of force scales times a geometric factor, is 
\begin{equation}
\lambda = \frac{2 \pi}{3} \frac{\Pi_0}{F}\left(\frac{R}{l_H}\right)^4.
\end{equation}
\section{Asymptotic analysis}
\label{sec:Asymptotic}
\begin{figure}[htbp]
\centering
\includegraphics{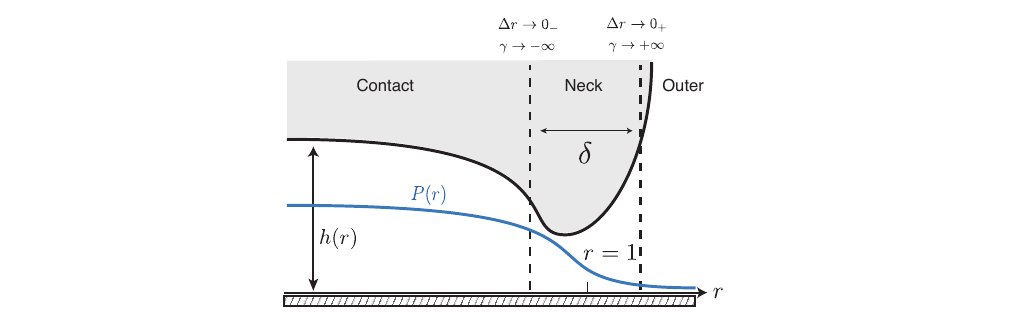}
  \caption{{\bf Asymptotic Analysis}. The underbelly of the soft solid, with example height $h(r)$ and pressure $P(r)$. We divide the soft solid into three regions: a contact region well underneath the solid, an outer region well outside, and a neck region of width $\delta(\lambda)$ connecting the two. The contact/neck interface is given by the limit $\Delta r:= r-1 \rightarrow 0_-$. In terms of the stretched variable $\gamma=\Delta r/\delta$, $\gamma\rightarrow -\infty$. The outer/neck edge is given by the limit $\Delta r \rightarrow 0_+$, $\gamma\rightarrow +\infty$.}
\label{fig:SIAsymptotics}
\end{figure}
{\bf Note:} In this section only, we omit tildes from variables: $r$, $h$, $P$, etc. are assumed non-dimensionalized by their Hertzian scales.

To analyze the properties of the height $h(r)$ and pressure $P(r)$ as $\lambda\rightarrow 0$, we employ an asymptotic matching approach~\cite{Bissett1989,sobac_leidenfrost_2014}. We divide the bottom of the soft solid into three regions: a contact region well under the solid, an outer region outside the solid, and a neck region connecting the two. These three regions are in shown in Fig.~\ref{fig:SIAsymptotics}. The contact and outer regions are naturally parameterized by the radius $r$. However the neck region, centered at $r=1$, has a width $\delta(\lambda)$, which narrows as $\lambda\rightarrow 0$. This observation motivates the definition of a stretched variable \begin{equation}
\gamma \equiv \frac{\Delta r}{\delta},
\label{eq:gamma}
\end{equation}
 where $\Delta r = r-1$. As $\delta(\lambda) \rightarrow 0$, $\gamma$ will remain an $O(1)$ variable parameterizing the neck.

As $\lambda\rightarrow0$, in the contact region the pressure profile is a perturbation of the Hertzian pressure profile. This perturbation will vanish as some power of $\lambda$. As the contact pressure tends to the Hertzian limit, the contact height profile will tend to zero, again vanishing as some power of $\lambda$. Applying the same logic in the outer region, as $\lambda \rightarrow 0$ the height profile will be a perturbation to the Hertzian height profile, and the pressure profile will approach zero. 

By analysing the Reynolds lubrication equation~\eqref{eq:SystemNonDim} in each region, we will patch the right edge of contact solution to the left edge of the neck solution, i.e., we will match as $\Delta r \rightarrow 0_- $ ($\gamma \rightarrow -\infty$), see Fig.~\ref{fig:SIAsymptotics}. Similarly, we will patch the left edge of the outer solution to the right edge of the neck solution, $\Delta r \rightarrow 0_+ $ ($\gamma \rightarrow +\infty$).  This patching process will determine the neck width $\delta(\lambda)$, and the asymptotic scaling of heights and pressures with $\lambda$. We begin our analysis with the contact region.
 
\subsection{Contact region}
We expand the pressure and height profiles as
\begin{equation}
  \begin{aligned}
    P= P_c(r)+ o(\lambda^0), \\
    h=\phi_c(\lambda)h_c(r) + o(\phi_c) ,
  \end{aligned}
  \label{eq:ContactExpansion}
\end{equation}
where 
\begin{equation}
P_c(r)= \sqrt{1-r^2}
\label{eq:HertzianPressure}
\end{equation}
is the Hertzian pressure profile, and $\phi_c(\lambda)$ is some power of $\lambda$, which captures the height profile's approach to zero. The subscript `c' stands for contact. With the $\lambda$ dependence peeled away, $h_c(r)$ is an $O(1)$ function which we initially assume to be unknown. 

Substituting the expansion Eq.~\eqref{eq:ContactExpansion} into the lubrication equation~\eqref{eq:SystemNonDim} yields
\begin{equation}
    \frac{1}{r}\frac{d}{dr}\left(r\frac{ h_c^3}{12}\frac{d}{dr}  \left(\frac{r}{\sqrt{1-r^2}}\right) \right) -\frac{\lambda\phi_c^{-4}}{h_c}=0.
    \label{eq:phi_c}
\end{equation}
Equation~\eqref{eq:phi_c} tells us the leading behavior of the height scaling in the contact region is given by $\phi_c(\lambda)=\lambda^{1/4}$. With this substitution, Eq.~\eqref{eq:phi_c} is an example of a Bernoulli equation. Although nonlinear, it may be transformed into a linear equation using the substitution $v(r)=h_c^4(r)$, to obtain the first-order ordinary differential equation 
\begin{equation}
\frac{d v}{dr} + \frac{4(2-r^2)}{3r(1-r^2)}v=\frac{16(1-r^2)^{1/2}}{r}.
\label{eq:v}
\end{equation}
Equation~\eqref{eq:v} is solved in terms of the ordinary hypergeometric function ${}_2F_1\left(a,b,c;z \right)$~\cite{riley1999mathematical,Abramowitz1964}. The constant of integration in Eq.~\eqref{eq:v} is set to $0$ by the requirement that $h_c(r)$ [and hence $v(r)$] must be finite at the origin, and we have the solution
\begin{equation}
h_c(r)=\left[ 6 (1-r^2)^{2/3} {}_2F_1\left(\frac{1}{6},\frac{4}{3},\frac{7}{3};r^2 \right) \right]^{1/4}.
\label{eq:h_c}
\end{equation}
To focus on the behavior of $h_c(r)$ as $r \rightarrow 1$, we now expand the hypergeometric function in Eq.~\eqref{eq:h_c} as $r \rightarrow 1$. To do so, we
first interchange the $r\rightarrow 1$ limit with an $r\rightarrow 0$ limit using a general identity for hypergeometric functions~\cite{Abramowitz1964}:
\begin{equation}
\begin{aligned}
\, _2F_1(a,b;c;z)&=\frac{\Gamma (c) (1-z)^{-a-b+c} \Gamma (a+b-c) \, _2F_1(c-a,c-b;-a-b+c+1;1-z)}{\Gamma (a) \Gamma (b)} \\
&+\frac{\Gamma (c) \Gamma (-a-b+c) \, _2F_1(a,b;a+b-c+1;1-z)}{\Gamma (c-a) \Gamma (c-b)},
\end{aligned}
\end{equation}
where $\Gamma(x)$ is the gamma function~\cite{riley1999mathematical}. Next, we use the series definition of ${}_2F_1(a,b;c;z)$ to expand about $r=0$. The result is that
\begin{align}
{}_2F_1\left(\frac{1}{6},\frac{4}{3},\frac{7}{3};r^2 \right) &= \frac{A}{r^{8/3}}+B(1-r^2)^{5/6}{}_2F_1\left(1,\frac{13}{6},\frac{11}{6};1-r^2 \right), \\
&=\frac{A}{r^{8/3}}+B(1-r^2)^{5/6}\left( 1+O(1-r) \right), \\
&=A+B(1-r^2)^{5/6}+O(1-r),
\end{align}
where $A= \Gamma(5/6)\Gamma(7/3)/\Gamma(13/6)\approx 1.24$, $B=\Gamma(-5/6)\Gamma(7/3)/\left[\Gamma(1/6)\Gamma(4/3)\right]=-1.6$ are known constants. Finally, we obtain the limiting behavior of $h_c$ as
\begin{equation}
h_c(r)=6^{1/4} (1-r^2)^{1/6} \left( A+B(1-r^2)^{5/6}+O(1-r)\right)^{1/4}.
\label{eq:hc}
\end{equation}
Expanding Eqs.~\eqref{eq:ContactExpansion} and~\eqref{eq:hc} in $\Delta r$ as $\Delta r \rightarrow 0_-$, we have our desired expansions in the contact region:
\begin{equation}
\begin{aligned}
P(\Delta r)&=(-2\Delta r)^{1/2} + O(\Delta r^{3/2}), \\
h(\Delta r)&=\phi_c(\lambda)\left[(6 A)^{1/4}(-2 \Delta r)^{1/6}\left(1 + \frac{B}{4A}(-2 \Delta r)^{5/6}+O(\Delta r) \right)\right],
\label{eq:LeadingContact}
\end{aligned}
\end{equation}
with
\begin{equation}
\phi_c(\lambda)=\lambda^{1/4}.
\end{equation}
Next, we turn to the outer region.
\subsection{Outer region}
As in the contact region, we expand the pressure and height profiles as
\begin{equation}
  \begin{aligned}
    P(r)&= \psi_o(\lambda) P_o(r)+o(\psi_o), \\
    h(r)&= h_o(r)+ o(\lambda^0) ,
  \end{aligned}
  \label{eq:OuterExpansion}
\end{equation}
 where the subscript `o' stands for outer. Here, the leading-order height profile is again given by the Hertzian solution~\cite{johnson_contact_1985},
\begin{equation}
h_o(r)=-1+\frac{r^2}{2}+\frac{1}{\pi}
\left[
\left(2-r^2\right)\arcsin\left({r}^{-1}\right)
+r\sqrt{1-r^{-2}}
\right].
\end{equation}
As $r\rightarrow 1_+$, $\Delta r \rightarrow 0_+$, we may expand $h_o(r)$ as
\begin{equation}
h_o(\Delta r)=\frac{8 \sqrt{2}}{3\pi}{\Delta r}^{3/2}+O(\Delta r^{5/2}).
\label{eq:LeadingOuter}
\end{equation}
In what follows we will not require the detailed form of $P_o(r)$. However, we note that upon substituting Eq.~\eqref{eq:OuterExpansion} into the Reynolds equation~\eqref{eq:SystemNonDim}, we immediately conclude that $\psi_o(\lambda)=\lambda$. The resulting linear differential equation, 
\begin{equation}
    \frac{d}{dr}\left(r\frac{ h_o^3}{12}\frac{dP_o(r)}{dr}  \right)= -\frac{r}{h_o},
\end{equation}
can be solved by integrating twice and applying the boundary conditions $P_o(r)=P'_o(r)=0$ as $r\rightarrow \infty$. We now proceed to the crucial matching conditions in the neck region.
\subsection{Neck region}
Now we perform the matching of the contact and outer solutions to the neck region. As before, we expand the height and pressure in the neck region: 
\begin{equation}
  \begin{aligned}
    P(r)&= \psi_n(\lambda) P_n(r)+o(\psi_n), \\
    h(r)&= \phi_n(\lambda) h_n(r)+o(\phi_n).
  \end{aligned}
\end{equation}
In this region, both the pressure and height tend to zero as $\lambda \rightarrow 0$, with powers $\psi_n$ and $\phi_n$ respectively. We read off $\psi_n$ and $\phi_n$ by expressing the contact solution, Eq.~\eqref{eq:LeadingContact}, and the outer solution, Eq.~\eqref{eq:LeadingOuter}, in terms of the stretched variable $\gamma$. Matching the contact solution Eq.~\eqref{eq:LeadingContact} as $\Delta r \rightarrow 0_-$ ($\gamma \rightarrow -\infty$), we have
\begin{equation}
\begin{aligned}
P(\gamma)&=(-2 \delta \gamma )^{1/2} + O((\delta \gamma) ^{3/2}), \\
h(\gamma)&=(6 A)^{1/4}\lambda^{1/4}(-2 \delta \gamma)^{1/6}\left(1 + \frac{B}{4A}(-2 \delta \gamma)^{5/6}+O(\delta \gamma) \right),
\end{aligned}
\end{equation}
from which we conclude
\begin{align}
\psi_n(\lambda)=\delta^{1/2}, \\
\phi_n(\lambda)=\lambda^{1/4}\delta^{1/6}.
\label{eq:phi_n_contact}
\end{align}
Matching the outer solution Eq.~\eqref{eq:LeadingOuter} as $\Delta r \rightarrow 0_+$ ($\gamma \rightarrow +\infty$), we have that
\begin{equation}
h(\gamma)=\frac{8 \sqrt{2}}{3 \pi}\left(\delta \gamma \right)^{3/2}+O\left((\delta \gamma)^{5/2}\right),
\end{equation}
from which we conclude that
\begin{equation}
\phi_n(\lambda)=\delta^{3/2}.
\label{eq:phi_n_outer}
\end{equation}
Equating our two expressions for $\phi_n(\lambda)$ from each edge of the neck region, Eqs.~\eqref{eq:phi_n_contact} and~\eqref{eq:phi_n_outer}, we find that
\begin{equation}
\delta(\lambda)=\lambda^{3/16}.
\label{eq:delta}
\end{equation}
With $\delta(\lambda)$ now fixed, we find expressions for the leading scalings of the height and pressure in the contact, outer, and neck regions. To summarize:
\begin{equation}
\begin{aligned}
\delta(\lambda)&=\lambda^{3/16},\\
\phi_c(\lambda)&=\lambda^{1/4},\\
\psi_o(\lambda)&=\lambda,\\
\phi_n(\lambda)&=\lambda^{9/32},\\
\psi_n(\lambda)&=\lambda^{3/32}.
\label{eq:Scalings}
\end{aligned}
\end{equation}
A subset of the expressions in Eq.~\eqref{eq:Scalings} are given as Eq.~(5) of the main text.

\section{Gap height scaling laws}
\label{sec:ScalingLaws}
To derive the scaling laws for gap height in the contact and neck regions given in Eq.~(1) and (6) of the main text, we re-dimensionalize the results of the asymptotic analysis, Eq.~\eqref{eq:Scalings}. For the contact region, we have that $\tilde{h} \sim \phi_c(\lambda)=\lambda^{1/4}$. In terms of dimensional variables, $h \sim \lambda^{1/4} (l_H^2/R)$. Using the definitions of $\lambda$ and $l_H$ in terms of materials parameters, we recover Eq.~(1) of the main text:
\begin{align}
h\sim \Pi_0^{1/4} \left( \frac{E}{1-\nu^2}\right)^{-1/3} R^{1/3} F^{1/12}.
\end{align}
Similarly, in the neck region, $\tilde{h} \sim \phi_n(\lambda)= \lambda^{9/32}$. In dimensional variables $h \sim \lambda^{9/32} (l_H^2/R)$, and we recover Eq.~(6) of the main text: 
\begin{equation}
h \sim \Pi_0^{9/32}\left(\frac{E}{1-\nu^2}\right)^{-7/24}F^{1/96}R^{5/12}.
\end{equation}

\section{Feasibility of future experimental validation}
\label{sec:Experiment}

In order to validate, or at a minimum establish consistency, with the theoretical results obtained here, experiments would require the following features. First, they must include samples in which at least one of the relevant physical properties (\textit{e.g.}~the radius, density, or Young's modulus) can be varied over an adequate range. Second, they must permit \textit{absolute} measurement of both the maximum height underneath the hydrogel and the neck height. Third, they must offer sufficient height resolution such that the predicted power laws can be established uniquely from other more common ones, or at least shown to be consistent with either.

\subsection{Sample Preparation}

Regarding hydrogel samples, the first and foremost requirement is that they have a water content of $\sim >$95\%.  Samples with a water content lower than this do not exhibit reliable floating behavior \cite{waitukaitis_coupling_2017}; hence the variety of possible hydrogels that can be used is quite limited.  Previous experiments used hydrogels meeting this criterion that were either commercially purchased, or synthesized in the laboratory. The commercially available spheres of Refs.~\cite{waitukaitis_bouncing_2018, waitukaitis_coupling_2017} have a fixed Young's modulus of approximately 50 kPa, and are available in a size range from a few millimeters to a couple of centimeters. Laboratory synthesized hydrogels, such as those used Ref.~\cite{pham_spontaneous_2017}, can be made with different Young's moduli, ranging from tens of kPa to order 100 kPa, but with lower water contents for larger Young's moduli. Hence, what is available to experiments currently is the possibility only to change the size of hydrogel spheres by approximately one order of magnitude in the range of a few mm to a few cm, and the possibility to change the Young's moduli by about a factor of 5 in the range 20-100 kPa. Because the water content must be approximately $\sim >$95\%, the density is necessarily fixed at approximately the density of water.

\subsection{Measurement Technique}

Regarding measurement techniques, arguably the most common technique for looking at floating Leidenfrost objects is lateral view videography, where a camera observes the floating object from the side and can view a gap underneath. This method was employed in Ref.~\cite{waitukaitis_coupling_2017} to track the visible gap height underneath floating hydrogels over \textit{very long} timescales. It is not sufficient to experimentally probe our results for several reasons. First, this method would only give access to the neck height, but not the maximum height underneath, which would be occluded via the neck. Second, the method only offers a spatial resolution of approximately 10 microns, whereas our models imply maximum/neck heights on the order of 10-20 microns for the hydrogel radii and Young's moduli that are experimentally possible. Hence the spatial resolution is insufficient. Related to this, the previous experiments of \cite{waitukaitis_coupling_2017} were not able to get any information during the first few seconds of floating due to this spatial issue, but also due to the requirement that the hydrogel has to be gently deposited onto the surface in order to float. What those experiments saw was the long-timescale irreversible shape changes due to evaporation of the hydrogel, \textit{not} the initial shape changes due to the balance between elasticity and vapor pressure that we see here. Hence, lateral view videography is certainly inadequate to investigate the models in this paper.

The next obvious candidate technique is interferometric imaging, as has been used to address floating liquid Leidenfrost droplets in Refs.~\cite{tran2012drop, burton_geometry_2012, veen_2012, bouillant_leidenfrost_2018}. In this technique, the underside of the floating object is illuminated by light passing through a transparent substrate. The interference pattern from reflections off the bottom of the object and the top of the substrate allow one to determine the profile of the underside of the object, yielding the possibility to measure both the maximum height and the neck height. When a single color of light is used for this technique, as in Refs.~\cite{burton_geometry_2012, bouillant_leidenfrost_2018}, the relative height profile underneath the object may be determined, but not the \textit{absolute height}. Hence interferometry of a single color is insufficient. With more colors of light, absolute height can be determined. For example, Ref.~\cite{veen_2012} used white light to measure the absolute underneath liquid droplets. However, this technique is limited in the range of absolute heights that can be measured by (a) the coherence length of the white light used, and (b) the level of confidence with which absolute height can unambiguously be recovered from the data analysis. Both of these limitations make white light interferometry insufficient for validating our models. For one, white light coherence lengths are only on the order of a few microns, whereas we expect absolute heights on the order of 10-20 microns for our experimentally accessible parameters. Secondly, the analysis techniques used to process previous data do not yield unambiguous results beyond a few microns.

While these two existing techniques are therefore insufficient to test our models, we can imagine an improvement to interferometric imaging that would allow us to do so. The main idea would be to use interferometry not with a white light source of short coherence length, but with multiple lasers of long coherence length. This would be significantly more complicated, as it requires precisely overlapping and aligning three separate lasers and then ultimately directing them onto three different cameras. Additionally, it would entail a height determination analysis that can unambiguously determine heights beyond the few microns of Refs.~\cite{veen_2012, tran2012drop}. Such a protocol, however, should be possible given (a) the large coherence lengths of the lasers and (b) the ability to lower hydrogels toward the surface at a constant speed, as in Ref.~\cite{waitukaitis_coupling_2017}. This would allow one to observe ``beats'' in the interference patterns of the lasers over tens or even hundreds of microns, thus reducing the ambiguity in the absolute height recovery analysis. 

\subsection{Quantifying Necessary Experimental Resolution}
\begin{figure}[htbp]
\centering
\includegraphics{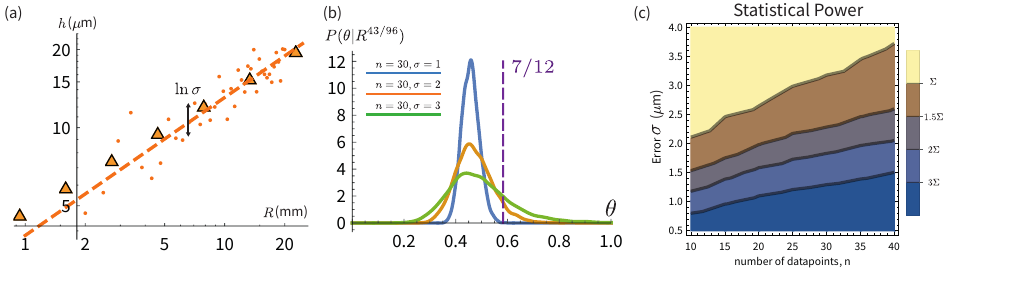}
  \caption{{\bf Quantifying height resolution to resolve neck and contact scaling}: We estimate the necessary measurement error $\sigma$ and dataset size $n$ for resolving the neck $h\sim R^{43/96}$ vs. contact $h\sim R^{7/12}$ scaling laws. (a)  An example synthetic dataset assuming the ground truth $h\sim R^{43/96}$. (b) Distributions $P(\theta|R^{43/96})$ on the best-fit slope $\theta$, with the contact scaling $\theta=7/12$ indicated. (c) How many standard deviations $\Sigma$ of $P(\theta|R^{43/96})$ one must take to observe a slope $\theta=7/12$, as function of $(n,\sigma)$.}
\label{fig:SIUQ}
\end{figure}

To quantify the necessary spatial resolution of such an interferometric setup, we now ask: how well must we determine gap height to resolve the difference between an $h\sim R^{7/12}$ contact height scaling versus an $h\sim R^{43/96}$ neck height scaling? In practice, resolving these scaling laws will be clearer for larger sphere radii. Here, we assume a range $R=\SI{2}{\milli\metre}-\SI{20}{\milli\metre}$, with all other parameters set to those of Ref.~\cite{waitukaitis_bouncing_2018}. This choice is consistent with commercially available hydrogels, as discussed above. Taking our simulated neck regime data as the ``ground truth" we generate $n$ synthetic data points using normally distributed errors with standard deviation $\sigma$. An example synthetic dataset in shown in Fig.~\ref{fig:SIUQ}(a). For each synthetic dataset, we generate a best-fit line, of slope $\theta$: $\theta$ is a point estimate for the ``ground truth" slope $R^{43/96}$. Repeating this process many ($\sim20,000$) times, we generate a distribution $P(\theta | R^{43/96})$ for the probability of observing $\theta$ given a ground truth $R^{43/96}$. Some example distributions are shown in Fig.~\ref{fig:SIUQ}(b). 

Using these distributions, we calculate how many standard deviations $\Sigma$ the contact scaling law $\theta=7/12$ lies away from the distribution mean: this gives a statistical power. Figure~\ref{fig:SIUQ}(c) shows a contour plot of statistical power in $(n, \sigma)$ space. For example, suppose we wish to resolve at a $3\Sigma$ level, and our apparatus has an experimental resolution of $\sigma=\SI{1}{\micro\metre}$: Fig.~\ref{fig:SIUQ}(c) indicates that we would require around $n=20$ data points. An experimental resolution of $\sigma=\SI{1}{\micro\metre}$ is reasonable within an interferometric setup, which gives resolution comparable to the wavelength(s) of 
 light used. We conclude that resolving the neck and contact scaling laws, which is one of the more delicate predictions our theory makes, is experimentally testable within a well-defined extension of previously used interferometric techniques.

\subsection{Comparing Contact and Neck Scaling in Simulation}
\begin{figure}[htbp]
\centering
\includegraphics{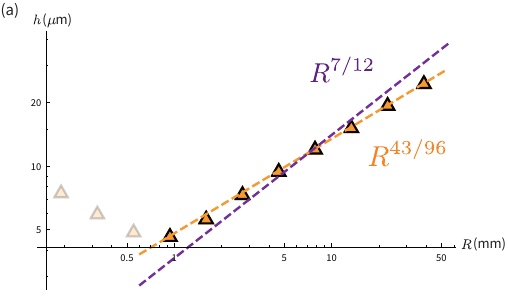}
  \caption{{\bf Comparing contact and neck scaling laws with numerical data}: Replotting the neck scaling data (orange triangles) shown in Fig. 2, alongside the predicted scalings of the contact ($h\sim R^{7/12}$) and neck ($h\sim R^{43/96}$) regions, emphasises the incompatibility of the contact scaling law with the neck data. A linear regression on neck data, removing the transition region (greyed out triangles), yields $h\sim R^{0.455 \pm 0.005}$ as a best-fit scaling law. The neck scaling law $43/96 \approx 0.448$ lies at the edge of the standard error here, whereas $7/12 \approx 0.583$ lies well outside this error.}
\label{fig:SIUQ2}
\end{figure}
We emphasise that our numerical data clearly resolves the existence of two scaling laws for the underbelly of the soft solid, independent of the question of experimental resolution. Figure~\ref{fig:SIUQ2} replots the neck scaling data of Fig. 2 in the main text, alongside the predicted contact and neck scaling laws. We see that the contact scaling, $h\sim R^{7/12}$, is incompatible with the neck data. Removing the transition region from the neck data, and fitting a simple linear regression, gives a best fit curve for the neck data of $h\sim R^{0.455\pm 0.005}$. This result is incompatible with the neck scaling, $7/12 \approx 0.583$. By contrast, our derived neck scaling law, $43/96\approx 0.448$, lies on the edge of the standard error here. We conclude that a second scaling law is indeed necessary to describe our observed data, and our asymptotic theory provides a result consistent with numerical data. In this sense, our asymptotic results provide a complete picture of the mechanism for elastic Leidenfrost floating: a dominant contact region with one scaling law $h\sim R^{7/12}$, and a narrow neck region with another scaling law $h \sim R^{43/96}$. 
\section{Simulation methodology}
\label{sec:Numerics}

Our hybrid/multiscale computational model is solved in COMSOL Multiphysics (COMSOL Ltd., Cambridge, UK; version 5.6) using the finite element method. Our approach is similar to that used by Chubynsky et al.~\cite{chubynsky_bouncing_2020} for the case of the isothermal impact of a droplet on a solid surface and by Chakraborty et al.~\cite{chakraborty_computational_2022} for the case of Leidenfrost droplets. The key difference is that the fluid dynamics problem inside the droplet is replaced by the solid mechanics problem in the present context of the elastic Leidenfrost effect.

We start with a soft elastic sphere of radius $R$ placed just above the heated surface. The initial distance from the surface should be similar to or somewhat larger than the expected equilibrium distance; too small distances create too large initial pressure forces on the sphere. Depending on materials parameters and $R$, initial distances $h_i$ between $10^{-3}R$ and $10^{-1}R$ were used. The soft hydrogel solid approaches the heated surface with an initial downward speed of $w=0.001$~m/s. To keep the initial approach slow, so that the equilibrium state is not badly overshot, gravity is initially off and is turned on when $t>h_i/w$.

As the basis of our simulations, we use the built-in \texttt{Solid Mechanics} module, which implements a finite-element solver for the equations of linear elasticity. Given the axial symmetry, the initial domain where these equations are solved is a half-disk with a semicircular boundary, and this is the domain which is filled with a finite element mesh. The shape of this domain evolves during the simulation, matching the shape of the soft solid. As vapor is treated within the lubrication approximation, meshing of the vapor domain is not required (see the details below).

The initially spherical surface of the hydrogel is divided into two hemispheres: the upper one where the free surface boundary condition is used (neglecting the influence of vapor), and the lower one where lubrication forces in the vapor film create a normal stress equal to the vapor pressure $P(r, t)$. 
It is convenient to map the lubrication equation~Eq. (2) for $P$ onto the lower part of the surface of the sphere and then solve it simultaneously with the linear elasticity equations in the bulk using the same mesh for both. In COMSOL this is straightforward to do by using the \texttt{Coefficient Form Boundary PDE} option. This is also facilitated by the fact that (as is indeed one of the requirements of the lubrication approximation) the surface of the soft solid is nearly horizontal everywhere the vapor pressure is significant; thus, the length coordinate $s$ along the contour of the hydrogel and the radial coordinate $r$ can be used interchangeably. To speed up convergence to equilibrium, we add Rayleigh damping (a standard option in COMSOL) with a mass damping parameter ($1.0$~s$^{-1}$) and a stiffness damping parameter (typically $10^{-4}$~s).

In our framework, the axisymmetric linear elasticity equations are solved for the dynamics of the hydrogel with the arbitrary Lagrangian-Eulerian approach employed for tracking the moving and deforming surface of the solid with high accuracy, whilst elements within the solid remain not too deformed. The solid domain is meshed using triangular elements with quadratic basis functions, with nodes of the mesh evolved using the Laplacian mesh smoothing technique. To resolve the narrow neck, which appears when $\lambda$ is small, the mesh is made finer near the bottom of the soft solid by using the \texttt{Size Expression} option in a manner similar to Ref.~\cite{chubynsky_bouncing_2020} (see the Supplemental Material of that reference). The resulting number of mesh elements is typically a few thousand. The time evolution is implemented using a second-order implicit backward differentiation formula (BDF2); the time step is adaptive with the maximum value typically $\Delta{t}=10^{-4}$~s. The resulting equations at each time step are solved using the multifrontal massively parallel sparse direct solver (MUMPS).

Convergence of the computed results was confirmed by repeating simulations with different mesh sizes and different maximum time steps.

\section{Quantifying finite size effects as $l_H/R \rightarrow 0$}
\label{sec:Deviations}
\begin{figure}[htbp]
\centering
\includegraphics{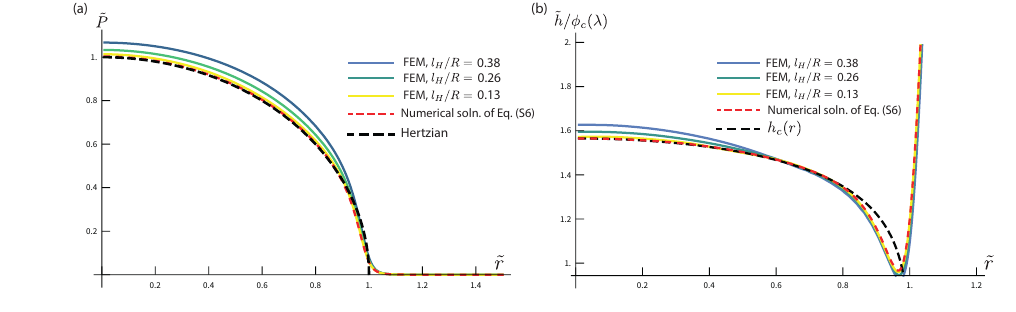}
  \caption{{\bf Quantifying finite-size effects as $l_H/R \rightarrow 0$:} Finite element pressure $\t P(r)$ (a) and rescaled height $\t h /\phi_c(\lambda)$ (b) profiles for fixed $\lambda=10^{-8}$, but varying $l_H/R$. For comparison, we show $\t P(r)$ and $\t h /\phi_c(\lambda)$ found via numerical solution of Eq.~\eqref{eq:SystemNonDim} (red dashed lines), and asymptotic results for the contact region (black dashed lines). As $l_H/R \rightarrow 0$, we see convergence between both simulation methods, with good agreement between numerics and our analytical result. Even for a non-negligible $l_H/R \approx 0.4$, finite size effects are restricted to a few percent across the solution. Data shown: For $l_H/R = 0.38$, $R =\qty{23.6}{\milli\metre}$ and $E =\qty{10}{\kilo\pascal}$. For $l_H/R = 0.26$, $R =\qty{38.7}{\milli\metre}$ and $E =\qty{50}{\kilo\pascal}$. For $l_H/R = 0.13$, $R =\qty{97.24}{\milli\metre}$ and $E=\qty{1000}{\kilo\pascal}$. All remaining materials parameters are as in Ref.~\cite{waitukaitis_bouncing_2018}, reproduced in Table~\ref{tbl:s2}.} 
\label{fig:SIDeviation}
\end{figure}

 In this section, we quantify the scale of deviations from our theory at finite values of $l_H/R$, stemming from the breakdown of Hertzian contact mechanics. In Fig.~\ref{fig:SIDeviation}, we show a series of height and pressure profiles obtained from our finite element method. Each profile has a different value of the sphere radius $R$ and Young's Modulus $E$, such that $\lambda$ is fixed to $\lambda=10^{-8}$, but $l_H/R$ varies (all other parameters are fixed to those of Ref.~\cite{waitukaitis_bouncing_2018} and Table~\ref{tbl:s2}). Fixing $\lambda$ allows us to focus solely on the finite size effects of varying $l_H/R$.

We compare our finite element results to two different calculations. Firstly, we compare to analytics. Our theory predicts that $\t P(r)$ asymptotes to its Hertzian value, Eq.~\eqref{eq:HertzianPressure}. In \S\ref{sec:Asymptotic}, we showed that the height $\t h(r)$ in the contact region asymptotes to $\t h(r)=\phi_c(\lambda) h_c(r)$, where  $\phi_c(\lambda)=\lambda^{1/4}$ is the height scaling of the contact region, Eq.~\eqref{eq:Scalings}, and $h_c(r)$ is the contact solution Eq.~\eqref{eq:h_c}. These predictions are shown in the black dashed curves of Fig.~\ref{fig:SIDeviation}. 

For a second comparison, we implement a direct numerical solution of Eq.~\eqref{eq:SystemNonDim}, which describes our coupled elasticity-fluid flow system. The numerical solutions for $\t P(r)$ and $\t h(r)$ found via this method are shown in the red dashed curves of Fig.~\ref{fig:SIDeviation}. To simulate Eq.~\eqref{eq:SystemNonDim}, we use the iterative scheme described in Ref.~\cite{hamrock_isothermal_1976}, in which trial values of $\t P(r)$, $\t h(r)$ are repeatedly substituted into the system Eq.~\eqref{eq:SystemNonDim} until convergence to a steady state is achieved. The derivation of Eq.~\eqref{eq:SystemNonDim} assumes the validity of Hertzian contact mechanics. As such, it represents the $l_H/R =0$ limit, which we can benchmark our finite element simulations against.  

Figure~\ref{fig:SIDeviation} shows that as $l_H/R \rightarrow 0$, the difference between our finite element simulations and the direct numerical solution of Eq.~\eqref{eq:SystemNonDim} vanish. Further, in the contact region there is good agreement between both methods and our analytical predictions  Eqs.~\eqref{eq:HertzianPressure},~\eqref{eq:h_c}. We also note that even for $l_H/R \approx 0.4$, deviations between simulation and theory are restricted to a few per cent. Taken together, these results demonstrate that our theory remains quantitatively accurate for the non-negligible values of $l_H/R$ which can occur in soft solids~\cite{waitukaitis_coupling_2017}.

\section{Cylindrical scaling laws}
\label{sec:Cylinder}
In the main text, we have focused on the case of a spherical soft solid. However, our approach may be applied more generally, and the initial geometry can make a dramatic impact on the resulting gap height. To illustrate this, here we consider the gap height scaling of an initially cylindrical geometry. Suppose the cylindrical axis is parallel to the $y$-axis (i.e., we have translational invariance along $y$) and let $l_y$ denote a unit length along this axis. $l_x$ denotes length along the $x$-axis, and is the analogue to the length scale $l$ found in the main text. As in the spherical case, we balance the integrated vapor pressure from the lubrication equation (where the characteristic length now is now $l_x$) with the total weight of the cylinder. We find
\begin{equation}
F\sim \left(\Pi_0 \frac{l_x^2}{h^4}\right) l_x l_y.
\end{equation}
In terms of a load per unit length, $W \equiv F/l_y$, we have
\begin{equation}
W\sim \Pi_0 \frac{l_x^3}{h^4}.
\label{eq:CylinderBalance}
\end{equation}
For a cylindrical contact, the Hertzian contact width is given by $l_x \sim \left[W R/E\right]^{1/2}$~\cite{johnson_contact_1985}, where $R$ is the cylindrical radius. Using this relation in Eq.~\eqref{eq:CylinderBalance} gives 
\begin{equation}
h \sim \Pi_0^{1/4} W^{1/8} E^{-3/8} R^{3/8}.
\end{equation}
Taking $W\sim R^2$ gives $h \sim R^{5/8}$, as stated in the main text.

\section{Materials parameters in Waitukaitis et al., Ref~\cite{waitukaitis_bouncing_2018}}
\label{sec:Materials}

\begin{table}[h!]
\label{tbl:s2}
  \begin{center}
    \caption{Material parameters found in Ref.\cite{waitukaitis_bouncing_2018}}
    \label{tab:Numbers}
    \begin{tabular}{c|c} 
      \hline 
      Materials Parameters \\
      \hline 
      Shear Viscosity $\eta$ &  \qty{2e{-5}}{\pascal\second}   \\
      Thermal Conductivity $\kappa$ & \qty{3e{-2}}{\watt\metre^{-1}\kelvin^{-1}} \\
      Latent Heat  $L$ & \qty{2.6e{6}}{\joule\kilogram^{-1}} \\
      Vapour Density  $\rho$ &\qty{5e{-1}}{\kilogram\metre^{-3}} \\
      Hydrogel Density  $\rho_s$ &\qty{10e{3}}{\kilogram\metre^{-3}} \\
      Temperature Difference  $\Delta T$ &\qty{115}{\kelvin} \\
      Young's Modulus $E$ &\qty{50e{3}}{\pascal}\\
      Poisson Ratio $\nu$ & $\approx 0.5 $\\
      \hline 
      Geometric Parameters  \\
      \hline 
      Sphere Radius  $R$ &\qty{7e{-3}}{\metre} \\
      \hline 
    \end{tabular}
  \end{center}
\end{table}
 For ease, here we reproduce the materials parameters of the hydrogel spheres used to obtain floating behavior in Ref.~\cite{waitukaitis_bouncing_2018}. These parameter values are used to obtain our estimate of $\lambda\sim 10^{-5}$ for the experimental setup of Ref.~\cite{waitukaitis_bouncing_2018}, as well as the data shown in Figs. 2, 3 of the main text.

\end{document}